%% file: Main-ActLocalizer.tex
\documentclass[journal]{vgtc}                     

\DeclareGraphicsExtensions{.png,PNG,.pdf,.PDF}

\onlineid{2463}

\vgtccategory{Research}

\title{Enhancing Single-Frame Supervision for\\Better Temporal Action Localization}

\author{%
  Changjian Chen,
  Jiashu Chen,
  Weikai Yang,
  Haoze Wang,
  Johannes Knittel, \\
  Xibin Zhao,
  Steffen Koch,
  Thomas Ertl,
  and Shixia Liu
}

\authorfooter{
  \item C.~Chen is with the College of Computer Science and Electronic Engineering, Hunan University. E-mail: changjianchen@hnu.edu.cn.
  \item
  	J.~Chen, W.~Yang, H.~Wang, X.~Zhao, and S.~Liu are with the School of Software, BNRist, Tsinghua University. S.~Liu is the corresponding author.
  	E-mail: \{\{cjs22, yangwk21, wang-hz22\}@mails., zxb, shixia@\}tsinghua.edu.cn.
   \item
       J.~Knittel, S.~Koch, and T.~Ertl are with University of Stuttgart.
        E-mail: {johannes.knittel, steffen.koch, Thomas.Ertl}@vis.uni-stuttgart.de.
}

\abstract{\input{0-abstract}}

\keywords{Temporal action localization, single-frame supervision, storyline visualization}

\teaser{
  \centering
  \includegraphics[width=\linewidth]{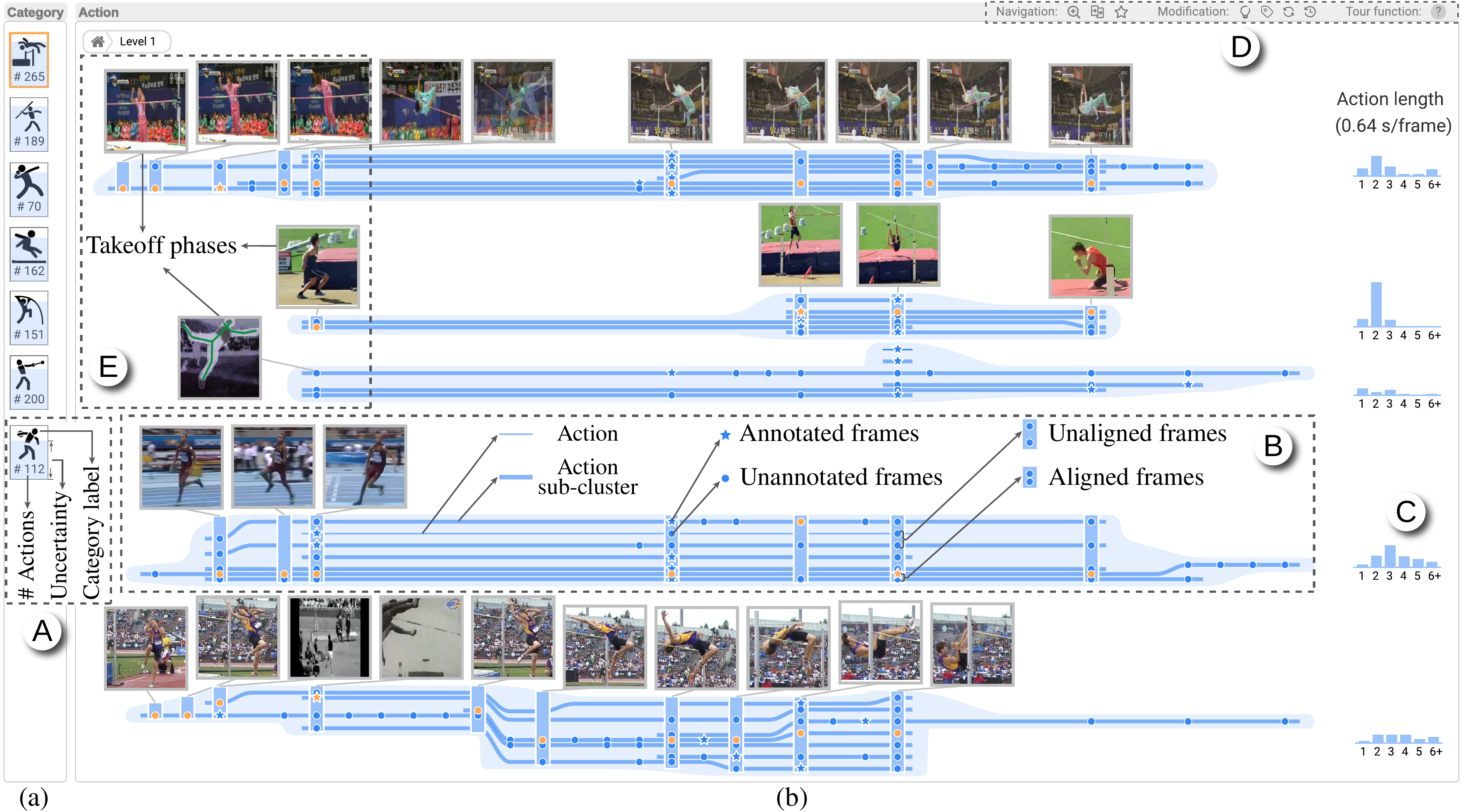}
     \vspace{-7mm}
  \caption{ActLocalizer: (a) a list to show action categories and their uncertainty; (b) a storyline to show actions and their alignments.}
  \label{fig:teaser}
     \vspace{-1mm}
}

\graphicspath{{figs/}{figures/}{pictures/}{images/}{./}} 

\usepackage{multirow}
\usepackage{color}
\usepackage{lipsum}                    
\usepackage{amssymb,amsmath}
\usepackage{wrapfig}
\usepackage{booktabs}
\usepackage{makecell}
\usepackage{dblfloatfix}
\usepackage[font=normal]{subcaption}

\usepackage{mathptmx}
\DeclareMathAlphabet{\mathcal}{OMS}{cmsy}{m}{n}

\newcommand{\tabincell}[2]{\begin{tabular}{@{}#1@{}}#2\end{tabular}}

\newcommand{\changjian}[1]{\textcolor{black}{#1}}

\def \etal {{\emph{et al}.\thinspace}}

\def \eg {{\emph{e.g}.\thinspace}}
\def \ie {{\emph{i.e}.\thinspace}}

\def \sys {ActLocalizer}

\def \Rexplore {\textbf{R1}}
\def \Rlocate {\textbf{R2}}
\def \Rprop {\textbf{R3}}
\def \Rstory {\textbf{R4}}

\def \Eone {$E_1$}
\def \Etwo {$E_2$}
\def \Ethree {$E_3$}
\def \Efour {$E_4$}

\newcommand{\myparagraph}[1]{\vspace{1mm}\noindent\textbf{#1}}

\def \mX {x}
\def \mY {y}

\def \mF {\mathbf{F}}
\def \mG {\mathbf{G}}

\def \mp {\mathrm{P}}
\def \mq {\mathrm{Q}}

\usepackage{mathptmx}                  

\begin{document}


\firstsection{Introduction}

\maketitle
\fontsize{9}{9} 
\input{1-introduction.tex}

\input{2-related.tex}

\input{3-system.tex}

\input{4-visualization.tex}

\input{5-experiments.tex}

\input{6-discussion.tex}

\input{7-conclusion.tex}

	

\bibliographystyle{abbrv-doi-hyperref}

\bibliography{reference}

\end{document}

%% file: 0-abstract.tex
\textit{Temporal action localization} aims to identify the boundaries and categories of actions in videos, such as scoring a goal in a football match.
Single-frame supervision has emerged as a \changjian{labor-efficient} way to train action localizers as it requires only one annotated frame per action.
However, it often suffers from poor performance due to the lack of precise boundary annotations.
To address this issue, we propose a visual analysis method that aligns similar actions and then propagates a few user-provided annotations (\eg, boundaries, category labels) to \changjian{similar} actions via the generated alignments.
Our method models the alignment between actions as a \changjian{heaviest path} problem and the annotation propagation as a quadratic optimization problem.
As the automatically generated alignments may not accurately match the associated actions and could produce inaccurate localization results, we develop a storyline visualization to explain the localization results of actions and their alignments.
This visualization facilitates users in correcting wrong localization results and misalignments.
The corrections are then used to improve the localization results of other actions.
The effectiveness of our method in improving localization performance is demonstrated through quantitative evaluation and a case study.


%% file: 1-introduction.tex

\begin{figure}[t]
\centering
\includegraphics[width=\linewidth]{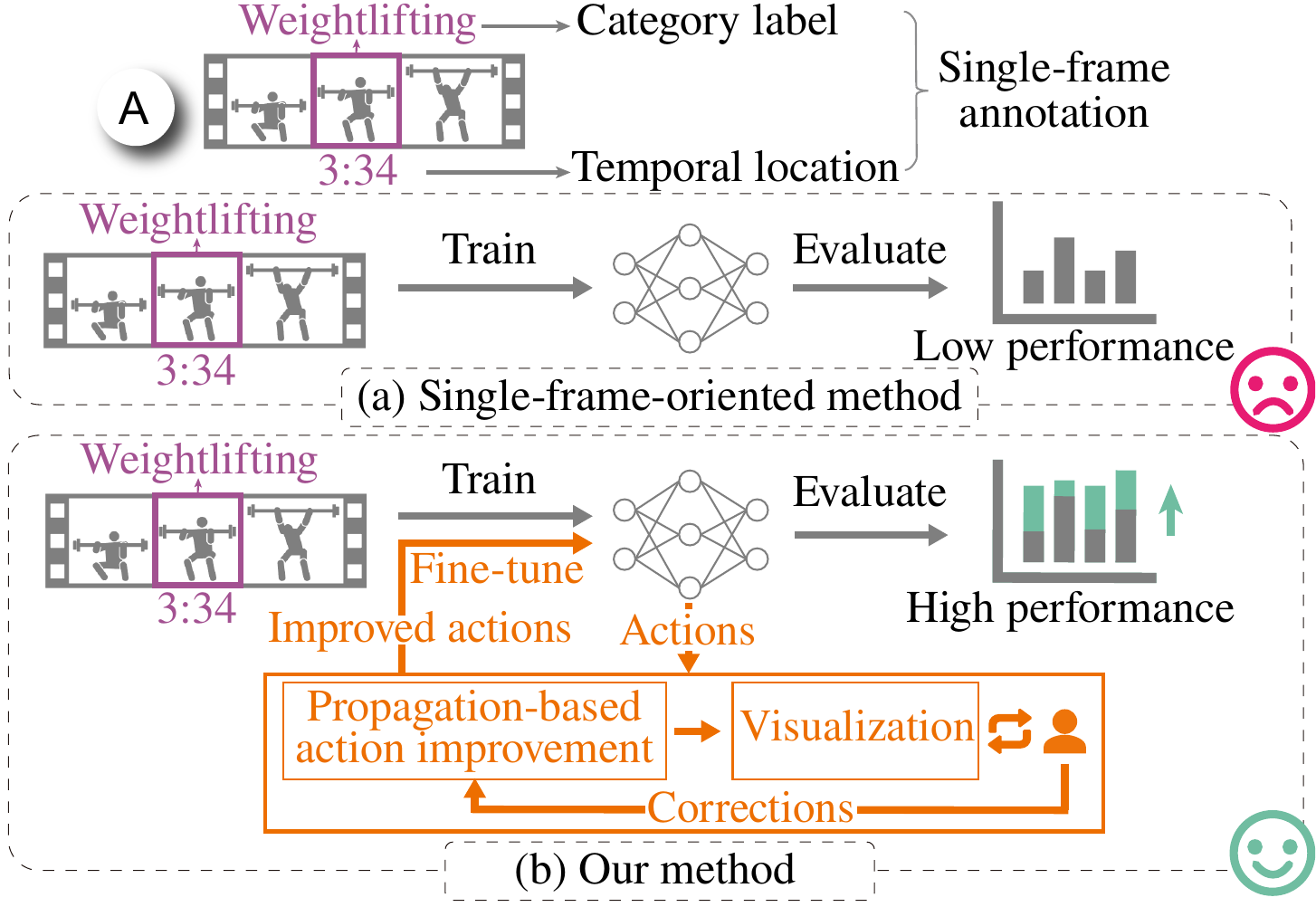}
\vspace{-5mm}
\caption{
The comparison between (a) the single-frame-oriented method and (b) the proposed visual analysis method (in {\textcolor{orange}{orange}}).}
\label{fig:motivation}
\vspace{-5mm}
\end{figure}

\emph{Temporal actions} refer to human motions or human-object interactions that occur in a video, such as scoring a goal in a football match and throwing a javelin~\cite{sultani2018real, xia2020survey}.
Detecting and analyzing temporal actions is important for a variety of applications, ranging from security surveillance to video moderation and home care~\cite{ma2020sf,zhao2022equivalent,yang2021background}.
As a result, \emph{temporal action localization} has emerged as a significant research topic in computer vision~\cite{xia2020survey}. 
It determines the boundaries and categories of actions in videos.
In recent years, deep learning models~\cite{zhang2022actionformer,liu2022end} have significantly improved the performance of temporal action localization by exploiting a large number of fully-annotated videos.
However, it is time-consuming to obtain these annotations since annotators need to seek back and forth through the videos to identify the precise boundaries of actions.
To address this issue, single-frame-oriented temporal action localization methods have been developed, which utilize \emph{single-frame annotations} to train action localizers~\cite{li2021temporal, ma2020sf}  (Fig.~\ref{fig:motivation}(a)).
A single-frame annotation of an action consists of the temporal location of one single frame (anchor frame) and its category label (Fig.~\ref{fig:motivation}A).
Annotators can provide such annotations by watching the video once with some extra pauses, which largely reduces annotation costs~\cite{ma2020sf}.
However, the performance of these methods is generally poor due to the lack of precise boundary annotations and the presence of noisy category labels.
\looseness=-1

An effective way to boost performance is to integrate humans into the automatic localization process.
First, a few imprecise boundaries and noisy category labels of detected actions are corrected by users.
Then, the corrections are propagated to \changjian{similar} actions \changjian{that share common human motions or human-object interactions} to improve the localization results.
\changjian{These results} are used to fine-tune the localizer for better performance.
This largely saves human efforts.
However, in the context of action localization, two challenges still persist.
First, as the dataset grows in size, manually identifying imprecise boundaries and noisy category labels becomes more challenging.
Second, due to the need to consider both frame similarities and inherent temporal relationships between sequential video frames, propagating corrections across \changjian{similar} actions presents a complex task.

To address these challenges, we develop {\sys}, a visual analysis method to 1) help users explore the localization results of actions for correction; 
2) propagate the corrections to \changjian{similar} actions through the generated alignments to improve localization performance. 
As shown in Fig.~\ref{fig:motivation}(b), {\sys} first trains an initial action localizer using the single-frame annotations to detect actions.
Then a propagation-based action improvement method is developed to improve the actions. 
Specifically, \changjian{similar} actions are aligned together by considering both similarities and temporal relationships between frames,
\changjian{which} is formulated as a \changjian{heaviest path} problem.
Based on the generated alignments, the localization results are improved by propagating the single-frame annotations to \changjian{similar} actions.
This is achieved by solving a quadratic optimization problem.
As the automatically generated alignments may not accurately match the associated actions and could produce inaccurate localization results,
we develop a storyline visualization to explain the actions and their alignments.
Users can interactively correct misalignments and wrong localization results.
The corrected alignments are incorporated into the \changjian{heaviest path} problem to generate better alignments, 
and the corrected localization results are propagated through the updated alignments to obtain improved actions.
Based on the improved actions, the action localizer is fine-tuned for better performance.
The effectiveness of the propagation-based action improvement method and the storyline is demonstrated through quantitative evaluation and a case study.
The source codes are available at: \href{http://actlocalizer.thuvis.org/}{\textcolor{black}{http://actlocalizer.thuvis.org/}}.

In summary, our contributions include:

\begin{itemize}[nosep]

\item\noindent{\textbf{A visual analysis tool} for iteratively improving the performance of action localizers through minimal user corrections on action annotations and alignments.
}

\item\noindent{\textbf{A storyline visualization} for exploring actions, examining alignments, and making necessary corrections.}

\item\noindent{\textbf{A propagation-based action improvement method} that effectively propagates user corrections to improve action localization results while saving human efforts.}

\end{itemize}

%% file: 2-related.tex
\section{Related Work}
\label{sec:related-work}


\subsection{Semi-Supervised Temporal Action Localization}
\label{subsec:related-work-learning}

Semi-supervised temporal action localization methods can be classified into two groups: boundary-oriented and single-frame-oriented methods.
\looseness=-1

Boundary-oriented methods utilize both annotated and unannotated videos to train action localizers~\cite{ji2019learning, wang2021self}.
\changjian{They introduce a consistency constraint between unannotated videos and their perturbed counterparts to improve the robustness against perturbations.}
For example, Ji~\etal~\cite{ji2019learning} kept the localization results unchanged for unannotated videos when the frames were resampled or some frames were masked. 
Recently, single-frame-oriented methods have been proposed to reduce annotation costs~\cite{li2021temporal, ma2020sf, yang2021background}. 
Compared to boundary-oriented methods, they achieve better performance with the same number of annotated frames. 
Compared to boundary-oriented methods, they \changjian{perform better when utilizing} the same number of annotated frames. 
Their training consists of two phrases.
First, an action localizer trained on the single-frame annotations is utilized to detect actions.
Second, these actions, which may be imprecise, are used to fine-tune the action localizer.
Assuming no irrelevant background frames in videos,
Li~\etal~\cite{li2021temporal} identified boundaries \changjian{by detecting} action changes between consecutive anchor frames.
Since the assumption is not always true,
this method may mispredict a background frame as an action frame.
To address this issue, Ma~\etal~\cite{ma2020sf} developed SF-Net, which determined boundaries by extending anchor frames to their temporally adjacent frames with high prediction confidence.

Despite the capabilities of single-frame-oriented methods, they can benefit from extra boundary annotations.
\changjian{Thus,}
we introduce a propagation-based method that improves the localization results of any single-frame-oriented method using these extra boundary annotations.

\subsection{Interactive Annotation for Sequence Data}

Interactive annotation methods for sequence data can be divided into two groups:
direct verification and iterative verification.

Direct verification methods utilize automatic algorithms to generate annotations~\cite{chen2021augmenting, kurzhals2017visualeye, piazentin2019historytracker}.
Users only need to validate or correct the generated annotations, greatly saving their efforts. 
For example, Kurzhals~\etal~\cite{kurzhals2017visualeye} utilized a video segmentation algorithm to partition eye-tracking data into multiple segments and clustered them.
\changjian{Users can annotate} multiple segments of a cluster in one go.
However, the accuracy of the automatically generated annotations significantly affects the annotation efficiency. 
To address this issue, iterative verification methods have been proposed~\cite{lekschas2020peax, tang2021videomoderator, yu2022pseudo, he2023videopro}.
In these methods, users annotated sequences that were recommended by an active learning model or identified in the visualization.
The annotations were then used to fine-tune the model for better performance.
This paradigm was also employed by Lekschas~\etal~\cite{lekschas2020peax} and Yu~\etal~\cite{yu2022pseudo} for identifying sequence data \changjian{of interest} and Tang~\etal~\cite{tang2021videomoderator} for annotating abnormal videos.
\looseness=-1

Among these methods, the most relevant one is VideoModerator~\cite{tang2021videomoderator}.
\changjian{It aims to} identify the videos with misinformation or offensive content.
Initially, a classifier was trained to recommend such videos,
The recommended videos were then analyzed and annotated in three coordinated views.
\changjian{These annotations were used to fine-tune the classifier. }
Since the classifier in VideoModerator cannot be utilized to detect the actions in videos, we have developed {\sys}, which effectively detects actions by combining single-frame annotations and a few user corrections (\eg, corrected boundaries).
Additionally, {\sys} utilized a storyline visualization \changjian{to aid users in analyzing and correcting actions, ultimately enhancing localization performance.}

\begin{figure*}[t]
        \vspace{-3mm}
  \centering
    \includegraphics[width=0.98\linewidth]{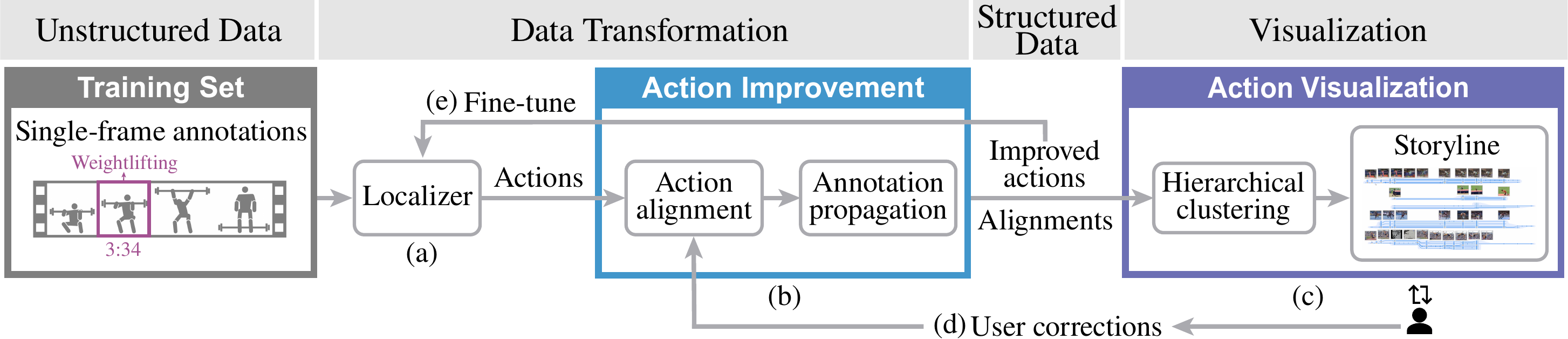}
            \caption{System overview: given videos and their single-frame annotations, (a) an action localizer is trained; (b) the action improvement module aligns \changjian{similar} actions together for propagating annotations; (c) the visualization module explains the actions and their alignments; 
            (d) users correct misalignments and wrong localization results;
            (e) the corrections are utilized to fine-tune the localizer. 
            } \looseness=-1
        \label{fig:analysis_pipeline}
        \vspace{-4mm}
\end{figure*}

\subsection{Video Visualization}
\label{subsec:related-work-vis}
Initial efforts on video visualization focus on visually illustrating the main content of a video by summarizing the representative frames~\cite{daniel2003video, kang2006space, nguyen2012video, tan2011imagehive, swift2022visualizing} or 
extracted attributes, including the motions of objects~\cite{chen2006visual, duffy2013glyph} and their trajectories~\cite{botchen2008action, hoeferlin2013interactive, meghdadi2013interactive}.
Recent efforts seek to combine machine learning techniques with interactive visualization for performing video analysis tasks,
including exploring the engagement of students in class~\cite{zeng2020emotioncues}, understanding inherent structures of movies~\cite{kurzhals2016visual, ma2020emotionmap}, 
and analyzing presentation techniques in TED talks~\cite{wu2020multimodal, zeng2019emoco}.
\looseness=-1

Orthogonal to these methods, our work focuses on improving the performance of action localizers.
For this purpose, we developed a storyline visualization that enables users to easily correct wrong localization results.
These corrections are then propagated by the propagation-based action improvement method to boost model performance.
\looseness=-1

%% file: 3-system.tex
\section{Requirement Analysis} 
\label{sec:requirement}

We worked closely with four machine learning experts ({\Eone}--{\Efour}) to develop {\sys}.
None of them are the co-authors of this paper.
\changjian{{\Eone} is a postdoctoral researcher, and {\Etwo} is a Ph.D. student.}
They have studied single-frame-oriented temporal action localization methods for over four years.
During their research studies, they found that the quality of the annotations limited the performance of the models.
Therefore, they wanted to interactively correct a few wrong localization results and propagate them to \changjian{similar} actions for better performance.
{\Ethree} and {\Efour} are two Ph.D. students who applied several single-frame-oriented methods to detect abnormal actions in surveillance videos for a project.
\changjian{These methods did not perform as expected, so they wanted to correct the localization results of some actions to enhance performance.}

We distilled the following requirements from four 40-70 minute semi-structured interviews with the experts and literature reviews.

\myparagraph{{\Rexplore} - Explore localization results and identify the wrong ones}.
\changjian{In practice, the experts had to examine each action individually to pinpoint the wrong localization results, including imprecise boundaries and noisy category labels. This becomes tedious when dealing with a large number of actions.}
To make the examination more \changjian{labor-efficient}, they desired an overview of the actions first.
Then, they wanted to examine the actions with imprecise boundaries and noisy labels at different levels of detail.
{\Etwo} mentioned, 
\changjian{``I prefer grouping similar actions together, each accompanied by a few representative frames to provide an overview.} The grouping results and the associated representative frames allow me to quickly identify the actions with wrong localization results that require further analysis.''

\myparagraph{{\Rlocate} - Correct wrong localization results more efficiently}.
Upon identifying wrong localization results, the experts required to correct them for better performance.
\changjian{While the experts can correct noisy category labels quickly,}
adjusting imprecise boundaries can be quite tedious.
For example, {\Eone} said, \changjian{``When refining the imprecise boundaries for one action, I find myself frequently rewinding the video to locate the precise boundaries, especially when they are unclear or ambiguous.''}
A more \changjian{labor-efficient} way is to recommend several boundary candidates for validation or refinement.
Additionally, {\Eone} suggested displaying boundary candidates in the context of neighboring frames. 
This would reduce the need for repeated video playback and thus save time and efforts.
\looseness=-1

\myparagraph{{\Rprop} - Propagate the the corrected localization results to \changjian{similar} actions}.
All the experts expressed the need to minimize the number of corrected localization results.
{\Ethree} commented, ``An \changjian{labor-efficient} way to achieve this is to propagate a few user-corrected localization results to \changjian{similar} actions.''
He noted that the propagation results depended on the alignments between actions.
Existing alignment methods only consider the similarities between the frames~\cite{chen2022towards, liu2018crowsourcing}
and ignore the temporal relationships between successive frames.
\changjian{This frequently leads to numerous misalignments}, such as aligning the upward and downward phases of two pull-up actions.
Thus, it is desired to align \changjian{similar} actions by considering both similarities and temporal relationships between frames.
\looseness=-1

\myparagraph{{\Rstory} - Correct the action alignments for better propagation}.
Since the alignments are generated automatically, some may not accurately match the associated actions~\cite{han2021video}.
These misalignments result in wrong propagation and subsequently degrade model performance.
Therefore, the experts wanted to examine the alignments between actions and understand how the corrected localization results are propagated through the alignments.
With a comprehensive understanding, they aimed to identify and correct a few misalignments.
\changjian{The remaining misalignments are expected to be corrected automatically for saving efforts.}
For example, {\Etwo} expressed the need for a tool to inspect action alignments, identify misalignments, and correct them for more effective propagation.
\looseness=-1

%% file: 4-visualization.tex
\section{Design of {\sys}}

\begin{figure*}[t]
  \centering
    \includegraphics[width=0.98\linewidth]{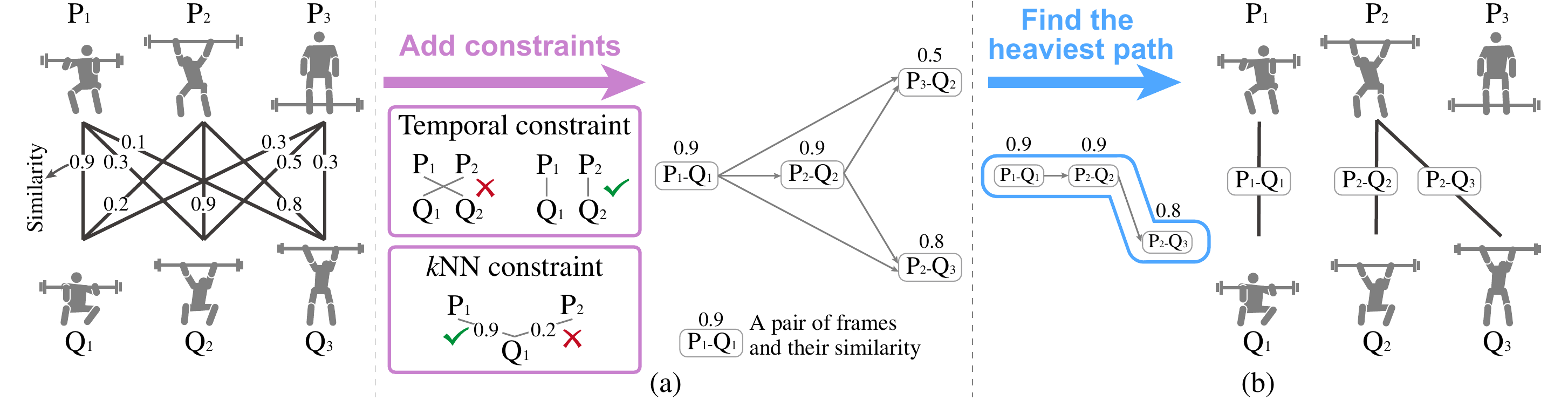}
            \caption{
                Action alignment: (a) adding constraints to construct a graph; (b) finding the \changjian{heaviest path} to obtain the alignments. 
        }
        \label{fig:alignment}
        \vspace{-3mm}
\end{figure*}

Guided by the requirements, we have developed {\sys} to iteratively improve the performance of action localizers. 
As shown in Fig.~\ref{fig:analysis_pipeline}, the analysis process begins with an input set of videos and their single-frame annotations.
Given the unstructured nature of this input, we first transform it into structured actions and alignments that are suitable for visualization.
The visualization then facilitates user feedback, which in turn refines the actions and alignments for further analysis.
Specifically, in the data transformation phase, an initial action localizer is trained on the input set for detecting the actions (Fig.~\ref{fig:analysis_pipeline}(a)).
The \textbf{action improvement module} (Fig.~\ref{fig:analysis_pipeline}(b)) first aligns \changjian{similar} actions together.
Then it propagates the single-frame annotations to the \changjian{similar} actions ({\Rprop}).
When transitioning to the visualization phase,
the \textbf{action visualization module} organizes the actions hierarchically and explains the actions and their alignments with a storyline ({\Rexplore},  Fig.~\ref{fig:analysis_pipeline}(c)).
During exploration, users can correct wrong localization results of actions ({\Rlocate}) and misalignments between them ({\Rstory}).
The corrected alignments are utilized to improve other alignments, and the corrected localization results are propagated through the updated alignments to obtain improved actions (Fig.~\ref{fig:analysis_pipeline}(d)).
Based on the improved actions, the action localizer is fine-tuned for better performance (Fig.~\ref{fig:analysis_pipeline}(e)).
\looseness=-1

\subsection{Propagation-based Action Improvement}

\label{sec:tal}

The propagation-based action improvement includes \changjian{two main components:} action alignment and annotation propagation.

\subsubsection{Action Alignment}
\label{sec:subsubalign}

Given two actions $\mathrm{P}$ and $\mathrm{Q}$, 
the action alignment establishes the temporal correspondence between their frames (Fig.~\ref{fig:alignment}).
A straightforward solution is to use dynamic time warping~\cite{muller2007dynamic}.
However, it requires \changjian{the alignments of each frame in one action with at least one frame in another action}.
In practice, detected actions usually \changjian{include} background frames, which should not be aligned with any action frames.
To address this issue, we adopted the method proposed by Tan~\etal~\cite{tan2009scalable}, which enables background frames to remain unaligned.
This method aligns actions by maximizing the total similarities of the aligned pairs:
\begin{equation}
\label{eq:alignment}
\begin{aligned}
\max_{\mathrm{z}} \sum_{i=1}^{|\mathrm{P}|}\sum_{j=1}^{|\mathrm{Q}|} \mathrm{z}_{ij} \mathrm{s}_{ij} \qquad \text{s.t.}\ \mathrm{z}_{ij} \in \{0,\ 1\}.
\end{aligned} 
\end{equation}
Here, $\mathrm{z}_{ij}$ is a binary variable indicating whether $\mathrm{P}_{i}$ and $\mathrm{Q}_{j}$ are aligned.
$\mathrm{P}_{i}$ and $\mathrm{Q}_{j}$ are the $i$-th and $j$-th frames of P and Q, respectively.
$\mathrm{s}_{ij}$ is their cosine similarity,
which is a common measure of frame similarity~\cite{cao2020few}.
$|\mathrm{P}|$ and $|\mathrm{Q}|$ are the numbers of frames in P and Q, respectively.
\looseness=-1

Since temporal relationships and frame similarities are the most important factors for aligning actions~\cite{tan2008accelerating}, we introduce the temporal and $k$NN constraints into Eq.~(\ref{eq:alignment}).
In addition, to help correct misalignments more effectively, 
the must-link/cannot-link constraints are also considered because they are easy for users to provide.
\changjian{Here we only consider the most common and domain-agnostic constraints. Other constraints, such as the interval constraint, can be easily integrated into our method.}
\looseness=-1

\begin{itemize}[nosep]

\item\noindent{\emph{Temporal constraint.}}
As the frames in an action are monotonic in time,
the aligned frame pairs should maintain such monotonicity
(\changjian{\ie}, $\forall (z_{ij}=z_{rt}=1) \wedge (i< r)\Rightarrow j\leq t$).

\item\noindent{\emph{$k$NN constraint.}}
The aligned frame pairs should have similar content as they represent the same phases of the corresponding actions.
Consequently, each frame within one action is \changjian{constrained to align with its $k$ nearest frames within another action}.
We choose $k$NN due to its simplicity and robustness~\cite{phdzhu2005semi, chen2021interactive}.
For each frame, $k$ is adaptively determined by the state-of-the-art method proposed by Zhao~\etal~\cite{zhao2021efficient}.
\changjian{$k$ of each frame} is adaptively determined by the state-of-the-art method proposed by Zhao~\etal~\cite{zhao2021efficient}.
It finds the smallest $k$ that yields a sufficiently large averaged prediction confidence for the $k$ nearest frames.

\item\noindent{\emph{Must-link/cannot-link constraints.}}
The must-link and cannot-link constraints are provided by users through interactions.
They constrain which frame pairs should be aligned and which should not.
\looseness=-1
\end{itemize}

\begin{figure}[b]
\vspace{-5mm}
\centering
\includegraphics[width=\linewidth]{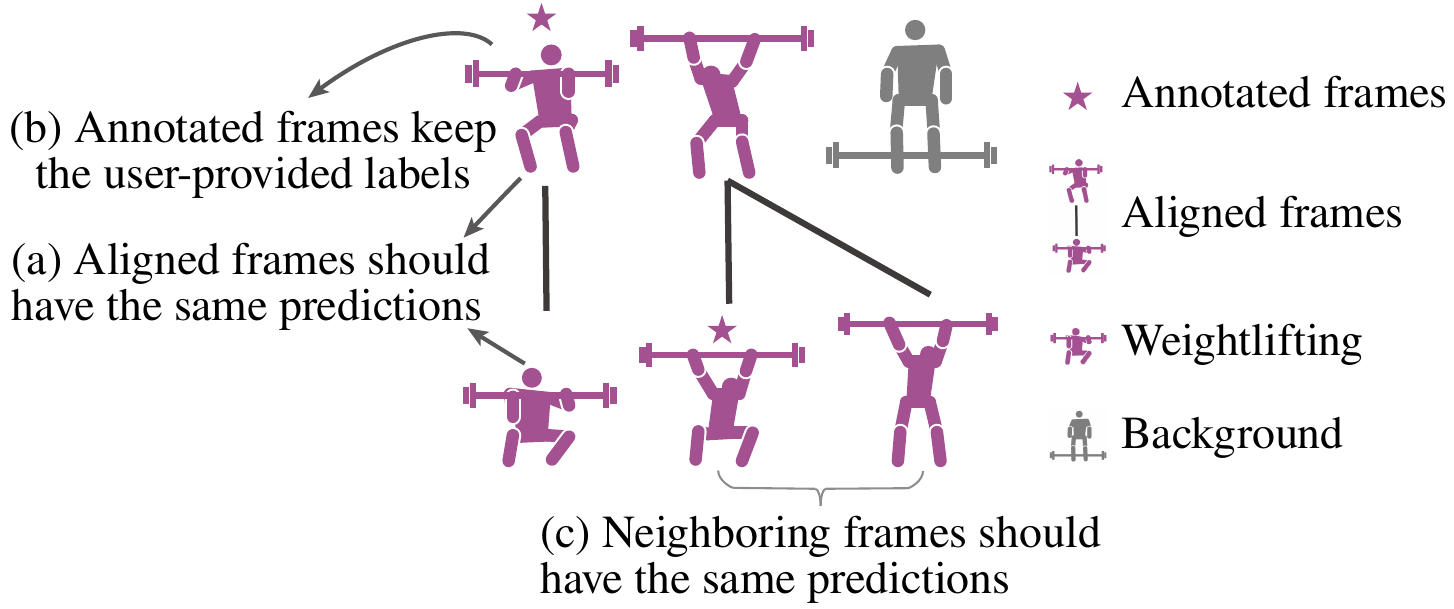}
    \vspace{-5mm}
    \caption{
  Annotation propagation:
  (a) aligned frames should have the same predictions; (b) annotated frames keep the user-provided labels; (c) neighboring frames should have the same predictions. \looseness=-1
    }
    \label{fig:propagation}
    \vspace{-2mm}
\end{figure}

All these constraints form a directed acyclic graph (Fig.~\ref{fig:alignment}(a)).
In the graph, \changjian{each node corresponds to a frame pair satisfying the $k$NN constraint, with its weight encoding the pair similarity.}
The edges between the nodes describe the temporal constraint.
For example, 
in Fig.~\ref{fig:alignment}(a), 
the edge from ``P$_1$-Q$_1$'' to ``P$_2$-Q$_2$''
is added because P$_1$ occurs before P$_2$ and Q$_1$ occurs before Q$_2$.
Each path of the graph represents a possible alignment result.
The must-link/cannot-link constraints specify the nodes to be included or excluded in the final path.
Given this graph, obtaining the optimal alignments (Eq.~(\ref{eq:alignment})) while satisfying the constraints is equivalent to finding a path that maximizes node weights.
This corresponds to a network \changjian{heaviest path} problem (Fig.~\ref{fig:alignment}(b)) and can be solved by dynamic programming.

\subsubsection{Annotation Propagation}
\label{subsubsec:propagation}
Annotation propagation improves localization results by propagating single-frame annotations and user corrections through the generated alignments \changjian{(Fig.~\ref{fig:propagation})}.
This propagation ensures that the aligned frames should have the same predictions (Fig.~\ref{fig:propagation}(a)) and the prediction of the annotated frames are consistent with their category labels (Fig.~\ref{fig:propagation}(b)).
If one boundary of an action is corrected by the user, the frames between its anchor frame and this boundary are also regarded as the annotated frames. 
In addition, due to the temporal relationships among frames, each unannotated frame should have the same predictions as its temporally closest annotated frame (Fig.~\ref{fig:propagation}(c)).
Accordingly, annotation propagation is formulated as a quadratic optimization problem:
\begin{equation}
\thinmuskip=-0.5mu
\thickmuskip=-2mu
\nulldelimiterspace=-1pt
\scriptspace=0pt
\label{eq:label-propagation}
    \min_{\mF} \ \ \sum_{i,j}^{T}{\mathrm{z}_{ij} \left\| \mF_i - \mF_j \right\|^2} \hspace{-0.2em}
  + \alpha \sum_{i}^{T} \delta_i \left\| \mF_i - \mG_i\right\|^2 + \beta \sum_{i}^{T} \min(\lVert\mF_i - \mF_i^{\mathrm{c}}\rVert^2, \tau).
\end{equation}
The first term minimizes the prediction difference between the aligned frames. 
The second term ensures the consistency between the predictions and the category labels for the annotated frames.
The third term minimizes the prediction difference between unannotated frames and their temporally closest annotated frames.
The weights $\alpha$ and $\beta$ balance the three terms.

Following the work of Iscen~\etal ~\cite{iscen2019label}, the \textbf{first term} utilizes the $L_2$ loss to measure the prediction difference between the aligned frames.
$T$ is the total number of frames.
$\mF_i$ represents the prediction of the $i$-th frame across different categories.
$\mathrm{z}_{ij} = 1$ indicates the $i$-th and $j$-th frames are aligned, and $0$ otherwise.
The \textbf{second term} also uses the $L_2$ loss to measure \changjian{the difference between the predictions of the annotated frames and their category labels.}
$\mG_i$ is the category label of the $i$-th frame.
$\delta_i=1$ indicates the $i$-th frame is annotated, and 0 otherwise.
The \textbf{third term} utilizes
the truncated mean squared error function~\cite{farha2019ms} to measure the prediction consistency between each unannotated frame $i$ and its temporally closest annotated frame $\mF_i^c$. 
The $i$-th frame is likely to be a background frame if the prediction difference between $\mF_i$ and $\mF_i^c$ is larger than a threshold $\tau$. 
Limiting the mean squared error to $\tau$ prevents the background frames from being mispredicted as action frames.
$\alpha$, $\beta$, and $\tau$ are determined by a grid search to balance the magnitude difference among the three terms.
The optimization problem Eq.~(\ref{eq:label-propagation}) can be solved by the gradient descent method~\cite{farha2019ms}.

\subsection{Action Visualization}

To help users explore and correct a large number of actions and their alignments, we cluster the actions into a hierarchy.
Based on the hierarchy, 
\changjian{we developed a storyline visualization} to explain the actions and their alignments.
Several interactions are also provided to help users correct misalignments and wrong localization results.

\subsubsection{Hierarchical Action Clustering}
\label{sec:hierarchy}
In {\sys}, an action hierarchy is built by a divisive method, which is one of the most widely used hierarchical clustering methods and \changjian{fast in computation}~\cite{reddy2018survey}.
The divisive method repeatedly applies a flat clustering algorithm to build the hierarchy in a top-down manner.
We choose K-medoids~\cite{park2009kmedoids} as the flat clustering algorithm due to its simplicity and robustness to noise~\cite{arora2016analysis,ma2018volumetric}.
Following the work of Wang~\etal~\cite{wang2012near}, the action similarities used for clustering are measured by the averaged similarities of aligned frame pairs between two actions.
\changjian{As there is no gold standard for determining the number of clusters~\cite{xu2005survey, togo2001food, newby2003dietary}, 
we employ the average silhouette width to evaluate the cluster results because it considers both the cluster compactness and separability,
and use a grid search for the best one.
Other methods for determining the number of clusters can also be used in our hierarchical action clustering method.}

To provide an overview of the action hierarchy, a set of \textbf{representative frames} ($\mathcal{U}$) is selected from each displayed cluster~\cite{chen2021oodanalyzer, chen2024unified}.
This selection aims to retain the representation quality while minimizing the number of the selected frames:
\begin{equation}
\label{eq:subset-selection}
\min_{\mathcal{U}} \sum_{j=1}^{T_c} \min_{i\in \mathcal{U}} d_{ij} + \gamma \lvert \mathcal{U}\rvert.
\end{equation}
The first term ensures better representativeness by minimizing the sum of the minimum dissimilarities between the selected frames and each unselected one.
The second term favors the selection of a small subset of frames.
$d_{ij} = 1 - s_{ij}$, 
where $s_{ij}$ is the cosine similarity between two frames.
$T_c$ is the total number of frames in an action cluster.
$|\mathcal{U}|$ is the number of the selected frames, and $\gamma$ is the weight to control the number of the selected frames.
According to the study of Elhamifar~\etal\cite{elhamifar2015dissimilarity}, it can be set as $\max_{ij} d_{ij} / \mathrm{M}$ to select around $\mathrm{M}$ frames.
In our implementation, $\mathrm{M}$ is set as 10 due to the space limit.
As optimizing Eq.~(\ref{eq:subset-selection}) is NP-hard, we employ a state-of-the-art approximate algorithm, the Alternating Direction Method of Multipliers~\cite{elhamifar2015dissimilarity, yang2022diagnosing}, 
to solve it.
This method decomposes the optimization problem into a set of simpler sub-problems and optimizes each sub-problem alternatively to obtain the final result.
\looseness=-1

\begin{figure}[b]
\vspace{-3mm}
\centering
\includegraphics[width=\linewidth]{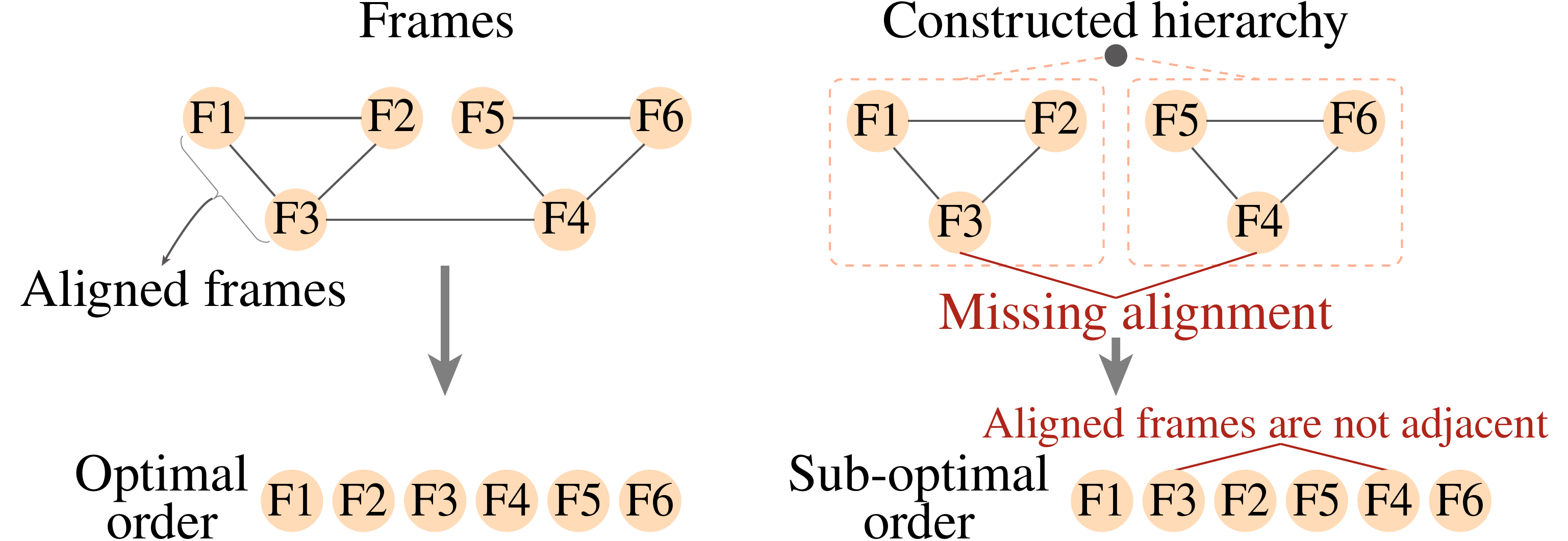}
    \caption{
    Some alignments are left out by the constructed hierarchy.
    }
    \label{fig:ordering}
    \vspace{-3mm}
\end{figure}

\begin{figure*}[t]
  \centering
    \includegraphics[width=0.98\linewidth]{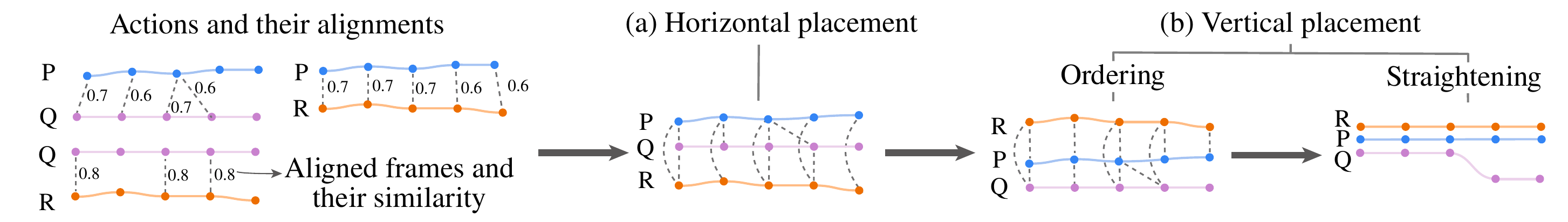}
        \caption{
                The storyline layout: (a) horizontal placement places aligned frames with higher similarities at the same horizontal positions; (b) vertical placement places aligned frames adjacently while minimizing line crossings (ordering) and the wiggle number (straightening).\looseness=-1}
        \label{fig:layout}
    \vspace{-3mm}
\end{figure*}

\subsubsection{Visual Design}
\changjian{Previous studies have demonstrated the effectiveness of the storyline metaphor in conveying sequential data and their temporal interconnections~\cite{liu2013storyflow,tanahashi2012design,zeng2020emotioncues}.}
Therefore, we adopt this metaphor to illustrate actions and their alignments.
The visual design consists of two parts: 
1) a category list to represent the action categories (Fig.~\ref{fig:teaser}(a)); 2) a storyline to illustrate the actions and their alignments (action view, Fig.~\ref{fig:teaser}(b)). 

In the category list, users examine the overview of the detected actions across different categories and their uncertainty. 
Each category is represented by a rectangle (Fig.~\ref{fig:teaser}A).
The simple drawing (\eg, $\vcenter{\hbox{\includegraphics[height=1.5\fontcharht\font`\B]{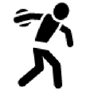}}}$) denotes the category label, and the number of actions in that category is shown below the drawing.
A solid filling style is utilized to encode the uncertainty of the category, which is inversely proportional to the average confidence of all the frames in this category. 
\changjian{The greater the filling height, the increased uncertainty.}
These categories are placed in descending order of uncertainty from top to bottom.

In the action view, the storyline enables users to explore actions at different levels of detail and identify the actions of interest.
As shown in Fig.~\ref{fig:teaser}B, each contour represents an action cluster, where each thin line ($\vcenter{\hbox{\includegraphics[height=1.5\fontcharht\font`\B]{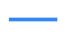}}}$) represents an action, and each thick line ($\vcenter{\hbox{\includegraphics[height=1.5\fontcharht\font`\B]{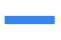}}}$) represents an action sub-cluster.
The circles ($\vcenter{\hbox{\includegraphics[height=1\fontcharht\font`\B]{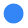}}}$) on each line represent the unannotated frames of the action, and the stars ($\vcenter{\hbox{\includegraphics[height=1\fontcharht\font`\B]{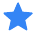}}}$) represent the annotated frames.
The horizontal positions of the frames represent their sequential orders.
The vertical distance between the circles/stars at the same horizontal positions indicates whether the associated frames are aligned.
The circles/stars are close if their associated frames are aligned; 
otherwise, they are unaligned.
The representative frames of a cluster are displayed above its contour with the associated circles/stars highlighted in orange.
The bar chart on the right side of each cluster shows the distribution of the action lengths (Fig.~\ref{fig:teaser}C).
\looseness=-1

\subsubsection{Layout}
\label{subsubsec:layout}
As the implementation of the category list is easy, here we focus on introducing how to generate the storyline.
The state-of-the-art solution, StoryFlow~\cite{liu2013storyflow}, generates legible storylines by placing entities in the same group adjacently while reducing line crossings, line wiggles, and white space.
Similar to StoryFlow, the aligned frames are expected to be placed adjacently while satisfying these legibility constraints.
\changjian{However, utilizing StoryFlow directly presents two key issues.}
First, 
StoryFlow assumes that each entity has a synchronized timestamp and
places the entities with the same timestamp at the same horizontal position.
As actions are usually captured at different times, their frames do not have such synchronized timestamps.
Second, StoryFlow assumes the existence of a location hierarchy for each timestamp.
Based on the hierarchies, StoryFlow reduces line crossings and wiggles while ensuring that entities in the same sub-hierarchy are placed adjacently along the vertical direction.
To apply StoryFlow to \changjian{generate our storyline}, we can build hierarchies for each timestamp based on the alignments between frames.
\changjian{However, as} some aligned frames may belong to different sub-hierarchies, such hierarchies may \changjian{overlook these} alignments and result in sub-optimal orders.
For example, the aligned pair ``F3-F4'' in Fig.~\ref{fig:ordering} is \changjian{overlooked} by the constructed hierarchy as they belong to two different sub-hierarchies.
This subsequently leads to a sub-optimal order where ``F3'' and ``F4'' are not placed adjacently.

To address these issues, we develop a layout method that consists of a horizontal placement and a vertical placement (Fig.~\ref{fig:layout}).
The horizontal placement aims to place aligned frames with higher similarities at the same horizontal positions (Fig.~\ref{fig:layout}(a)), and the vertical placement seeks to reduce line crossings, line wiggles, and white space (Fig.~\ref{fig:layout}(b)).
\looseness=-1

\begin{figure}[b]
\centering
\includegraphics[width=\linewidth]{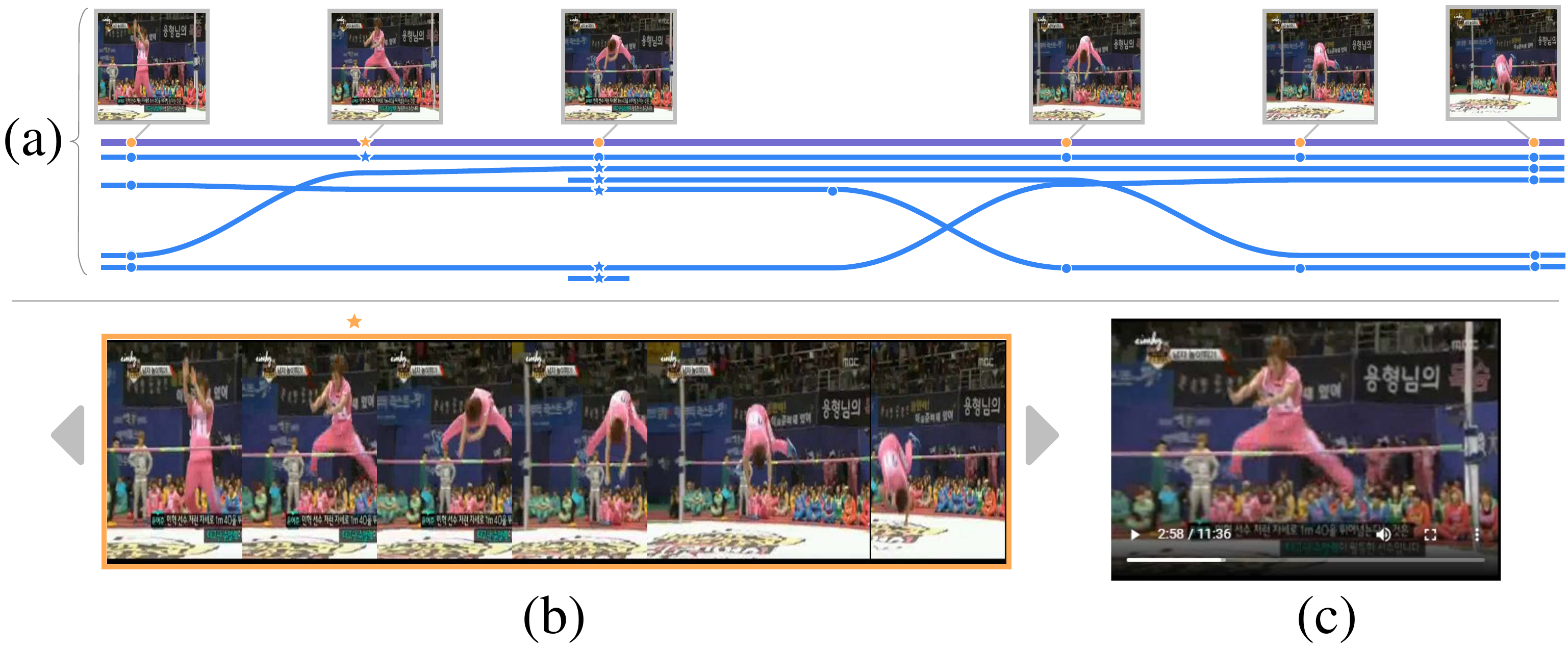}
    \caption{
        Selecting an action of interest: (a) the storyline to show the alignments between the selected action and its neighbors; (b) the content of the associated frames; (c) the associated video clip.
    }
    \label{fig:comparison}
\end{figure}

\myparagraph{Horizontal placement}.
Mathematically, the horizontal placement is expressed as:
\begin{equation}
\label{eq:horizontal-placement}
\begin{aligned}
& \max_{\mX} & & \underset{\mp, \mq \in \mathcal{A}\ }{\sum} \underset{ \mX(\mp_i) = \mX(\mq_j)}{\sum} (z_{ij} +\lambda \cdot \mathrm{s}_{ij}) \\
& \text{ s.t.} & & \mX(\mp_i) < \mX(\mp_j)\ \ \ \mathrm{if}\ \ \ i < j,\ \forall\ \mp \in \mathcal{A}.
\end{aligned}
\end{equation}
The first term ensures that aligned frames are placed at the same horizontal positions.
The second term encourages similar frames to be placed at the same horizontal positions.
The constraint \changjian{guarantees that the frames' horizontal positions} are consistent with their sequential orders.
$\mathcal{A}$ is a set of actions,
$\mX(\cdot)$ is the horizontal position of a frame,
$z_{ij}$ is a binary variable indicating whether frames $\mp_i$ and $\mq_j$ are aligned ($z_{ij} = 1$) or not ($z_{ij} = 0$),
$s_{ij}$ is their cosine similarity,
and $\lambda$ is the weight to balance the two terms.
Since the primary goal of this step is to place aligned frames at the same positions, the weight of the first term (alignment) should be larger than that of the second term (similarity).
Therefore, we set $\lambda$ as 0.1 in our implementation.
Following the work of Feng~\etal~\cite{feng1987progressive},
this optimization problem is solved by a greedy strategy.
This strategy iteratively determines the horizontal position of the frames in one action at a time, 
\changjian{while freezing the frames of previously placed actions.}
\looseness=-1

\myparagraph{Vertical placement}. 
The vertical placement consists of three steps: ordering, straightening, and compaction (Fig.~\ref{fig:layout}(b)). 
The compaction algorithm of StoryFlow is employed to reduce unnecessary white space, so we focus on introducing the first two steps. 
\looseness =-1

\textit{Ordering}.
This step ensures that the aligned frames are placed adjacently along the vertical direction while minimizing line crossings. 
\begin{equation}
\label{eq:ordering}
\begin{aligned}
\min_{\phi}  \underset{\mp, \mq \in \mathcal{A}\ }{\sum} \underset{\mX(\mp_i) = \mX(\mq_j)}{\sum} z_{ij} \cdot \mathbb{I}(|\phi(\mp_i) - \phi(\mq_j)| > 1) + \mu C(\phi)
\end{aligned}
\end{equation}
The first term ensures that aligned frames are placed in adjacent \changjian{vertical} positions,
while the second term penalizes line crossings.
Here, $\phi(\cdot)$ denotes the vertical order of a frame among all frames at the same horizontal position
and $\mathrm{C}(\phi)$ is the number of line crossings.
$\mu$ is the weight to balance the two terms, which is set as 3 in our implementation to \changjian{prioritize the reduction of line crossings}.
The global optimum of Eq.~(\ref{eq:ordering}) is obtained using state compression dynamic programming~\cite{held1962dynamic}. 

\textit{Straightening}.
The main goal of straightening is to minimize the number of line wiggles while also preserving the vertical distances between frames based on their alignments.
This is formulated as a constrained optimization problem:
\begin{equation}
\label{eq:straightening}
\begin{aligned}
& \min_{\mY} & & \underset{\mp \in \mathcal{A}\ }{\sum} \sum_{i}\mathbb{I}[\mY(\mp_i) \neq \mY(\mp_{i+1})] \\ 
& & & + \mu \underset{\mp, \mq \in \mathcal{A}\ }{\sum} 
\underset{\substack{\mX(\mp_i)=\mX(\mq_j), \\ |\phi(\mp_i) - \phi(\mq_j)| = 1}}
{\sum} [|\mY(\mp_i) - \mY(\mq_j)| - (1-z_{ij})]^2 \\
& \text{s.t.} & & \mY(\mp_i) < \mY(\mq_j)\ \ \ \mathrm{if}\ \ \  
\phi(\mp_i) < \phi(\mq_j).
\end{aligned}
\end{equation}
The first term penalizes the line wiggles,
while the second term \changjian{guarantees vertical closeness} between aligned frames and maintains a specific vertical distance between unaligned frames. 
$\mY(\cdot)$ is the vertical position of a frame.
The weight $\mu$ is used to balance the magnitude difference between the two terms and is determined by a grid search.
Eq.~(\ref{eq:straightening}) can be solved by dynamic programming.
\looseness=-1

\subsubsection{Interactive Exploration and Correction}
To help users easily correct misalignments and wrong localization results,
{\sys} provides two types of interactions: action filtering and interactive correction.

\myparagraph{Action filtering}.
When users identify an action of interest, 
they can filter this action and its neighbors to better analyze their alignments (Fig.~\ref{fig:comparison}).
The selected action and its neighbors are placed on the top (Fig.~\ref{fig:comparison}(a)).
The frames (Fig.~\ref{fig:comparison}(b)) and the associated video clip (Fig.~\ref{fig:comparison}(c)) of the selected action are also provided to facilitate the examination and analysis in context.

\myparagraph{Interactive correction}. 
{\sys} \changjian{allows users to enhance the performance of action localizers by} interactively correcting misalignments and wrong localization results.

\textit{Correcting misalignments}.
When users identify misalignments, they can correct them directly in the visualization.
Users can drag two frames closer to indicate that \changjian{they must be aligned} and farther apart to indicate that they cannot be aligned.
The corrections are converted to the must-link and cannot-link constraints in the alignment algorithm,
which are then utilized to update the other alignments between actions.

\textit{Correcting wrong localization results}.
If users identify wrong localization results, including imprecise boundaries and noisy category labels, 
they can right-click the frames to annotate the boundaries or click $\vcenter{\hbox{\includegraphics[height=1.5\fontcharht\font`\B]{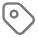}}}$ in Fig.~\ref{fig:teaser}D to change the category labels.
The corrected localization results are then propagated to other \changjian{similar} actions.
To reduce user efforts in \changjian{this} correction process, {\sys} allows them to annotate a rough boundary and recommends a more precise one based on the rough one. 
The recommended boundary is obtained by propagating the \changjian{annotated} rough boundary \changjian{using} the propagation method described in Sec.~\ref{subsubsec:propagation}.
If the recommended boundary is still imprecise, users can annotate a new one based on the recommendation and iteratively refine it.
\looseness=-1

%% file: 5-experiments.tex
\section{Evaluation}
\label{sec:learning}

We conducted quantitative evaluation to evaluate the performance of the propagation-based action improvement method and a case study to demonstrate the effectiveness of {\sys}.

\subsection{Quantitative Evaluation on Action Localization}

\myparagraph{Datasets}.
In this experiment, two widely used datasets are employed.
The first dataset, \textbf{THUMOS14}~\cite{idrees2017thumos}, contains 200 training videos (63,575 frames) and 213 test videos (70,044 frames) with 20 sports categories.
We noticed that some cliff diving actions were annotated as diving.
\changjian{To mitigate this inconsistency, we merged these two categories into a single category, namely "diving."}
The second dataset, \textbf{BEOID}~\cite{damen2014you}, contains 58 videos (6,588 frames) with 30 action categories.
They are recorded in six different scenes (\eg, kitchens and gyms).
The videos are randomly split into an $80\%$ training set and a $20\%$ test set.
Each action in the training sets of both datasets has a single-frame annotation~\cite{ma2020sf}.

\begin{table}[t]
    \vspace{1.5mm}
     \caption{Performance comparison between our method and SF-Net in terms of the mAP (in \%) on two benchmark datasets.}
    \begin{subtable}[h]{0.5\textwidth}
        \centering
          \begin{tabular}{c|c|c|c|c}
          \toprule
            Annotations & 2\%  & 5\% & 10\% & 20\% \\
          \hline
            \tabincell{c}{SF-Net \\ + our method} & \tabincell{c}{$\mathbf{42.70}$ \\ (+$\mathbf{1.10}$)} & \tabincell{c}{$\mathbf{42.91}$ \\ (+$\mathbf{1.18}$)} & \tabincell{c}{$\mathbf{43.08}$ \\ (+$\mathbf{1.04}$)} &  \tabincell{c}{$\mathbf{43.49}$ \\ (+$\mathbf{0.74}$)}  \\
          \hline
            \tabincell{c}{SF-Net}  & \tabincell{c}{$41.60$ } & \tabincell{c}{$41.73$} & \tabincell{c}{$42.04$}   & \tabincell{c}{$42.75$} \\
          \bottomrule
          \end{tabular} 
        \vspace{0mm}
      \caption{\textbf{THUMOS14}}
        \vspace{1.5mm}
      \label{tab:thumos}
    \end{subtable}
    \hfill
    \begin{subtable}[h]{0.5\textwidth}
        \centering
          \begin{tabular}{c|c|c|c|c}
          \toprule
            Annotations &  2\%  & 5\% & 10\% & 20\% \\
          \hline
            \tabincell{c}{SF-Net \\ + our method} &  \tabincell{c}{$\mathbf{32.51}$ \\ (+$\mathbf{1.03}$)} & \tabincell{c}{$\mathbf{32.79}$ \\ (+$\mathbf{1.18}$)} & \tabincell{c}{$\mathbf{33.21}$ \\ (+$\mathbf{1.25}$)} &  \tabincell{c}{$\mathbf{33.36}$ \\ (+$\mathbf{1.03}$)}  \\
          \hline
            \tabincell{c}{SF-Net}  & \tabincell{c}{$31.48$ } & \tabincell{c}{$31.61$} & \tabincell{c}{$31.96$}   & \tabincell{c}{$32.33$} \\
          \bottomrule
          \end{tabular} 
        \vspace{0mm}
        \caption{\textbf{BEOID}}
         \vspace{-1.5mm}
        \label{tab:beoid}
     \end{subtable}
    \hfill

  \vspace{-4mm}
     \label{tab:map}
\end{table}

\begin{figure*}[t]
    \vspace{-3mm}
  \centering
    \includegraphics[width=0.98\linewidth]{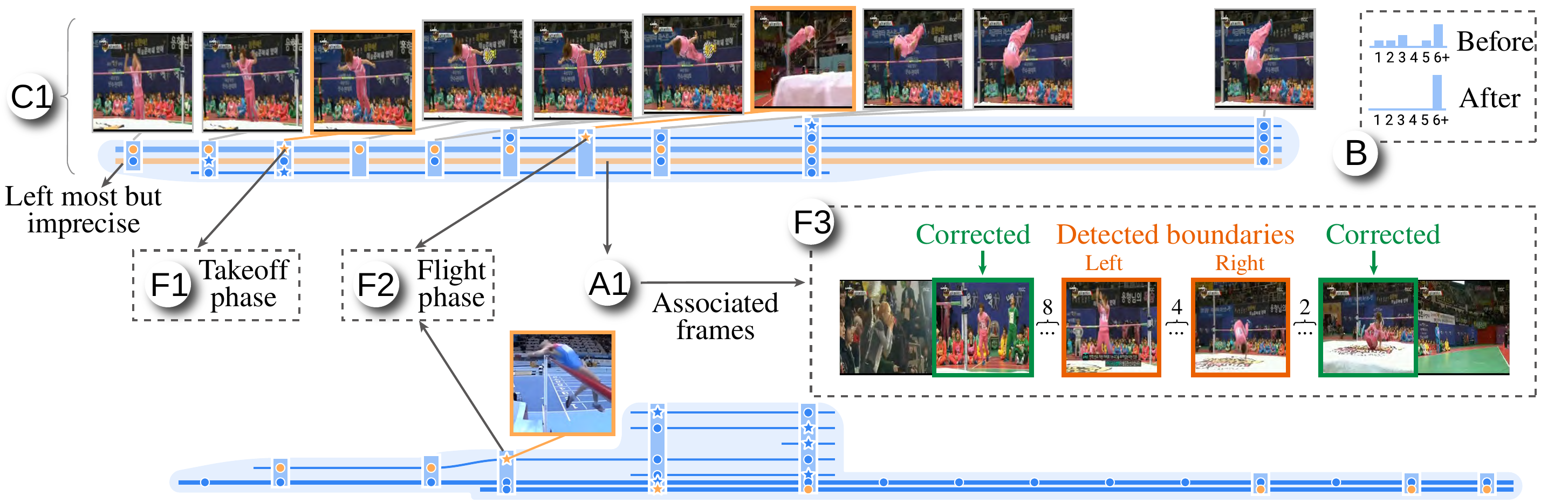}
    \caption{When analyzing the high jumping category, C1 with more frames on the left side and imprecise boundaries is identified.}
\label{fig:annotating}
    \vspace{-3mm}
\end{figure*}

\begin{figure*}[b]
    \vspace{-3mm}
  \centering
    \includegraphics[width=0.98\linewidth]{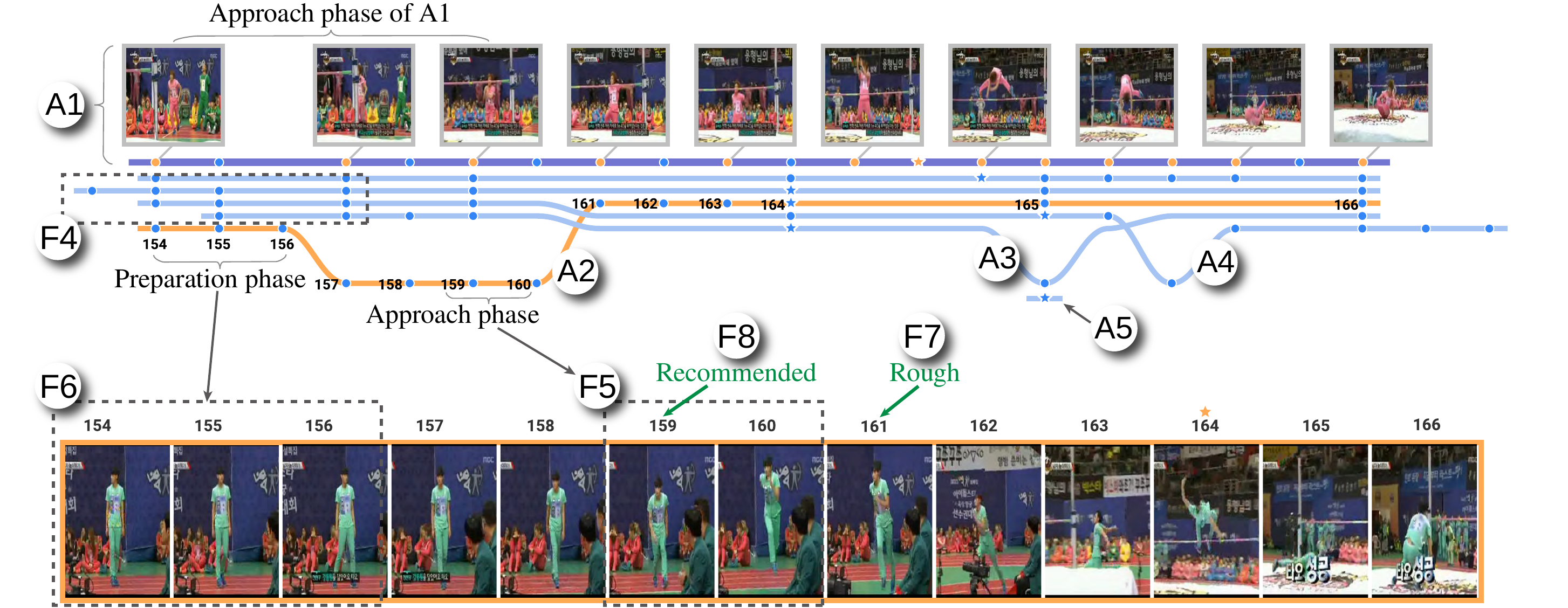}
    \caption{After correcting the imprecise boundaries of A1 in the high jumping category, more frames of the approach phase are detected. Some unaligned frames are also identified in A1's neighbors (A2--A5).}
     \vspace{-3mm}
    \label{fig:correcting-alignments}
\end{figure*}

\myparagraph{Experimental settings}.
Theoretically, our propagation-based action improvement method can enhance the performance of any single-frame-oriented localization method.
In this experiment, we selected the state-of-the-art one, SF-Net, as a representative example.
The effectiveness of our method is demonstrated by comparing the performance of SF-Net with and without our propagation-based action improvement method.
To conduct the experiments while \changjian{saving user annotation time and efforts}, we simulated user-provided boundary annotations by randomly sampling $v$\% ($v \in \{ 2,\ 5,\ 10,\ 20 \}$) of actions and using the ground truth boundaries as the annotations.
For SF-Net without our method, the RGB features extracted by Swin Transformer~\cite{liu2022video} and optical flow features extracted by I3D network~\cite{carreira2017quo} (a pretrained model) are utilized for training.
Based on the extracted features, SF-Net uses both single-frame and user-provided boundary annotations to train an action localizer, which is utilized to detect actions. 
Then, the detected actions are utilized to fine-tune the action localizer.
For SF-Net with our method, the proposed propagation-based action improvement method is added to improve the detected actions by propagating the user-provided boundary annotations to \changjian{similar} actions.

\myparagraph{Result}.
Performance was evaluated by mAP@IoU and mAP@Hit~\cite{ma2020sf}, the commonly used measures for temporal action localization.
We found that using the mAP@IoU and mAP@Hit scores led to similar conclusions. 
Therefore, we included only the mAP@IoU scores in the paper and abbreviate them as mAP scores, while the mAP@Hit scores can be accessed in the supplementary material.
As shown in Table~\ref{tab:map}, 
our method improved the localization performance on the two datasets.

\begin{figure*}[t]
    \vspace{-3mm}
  \centering
    \includegraphics[width=0.98\linewidth]{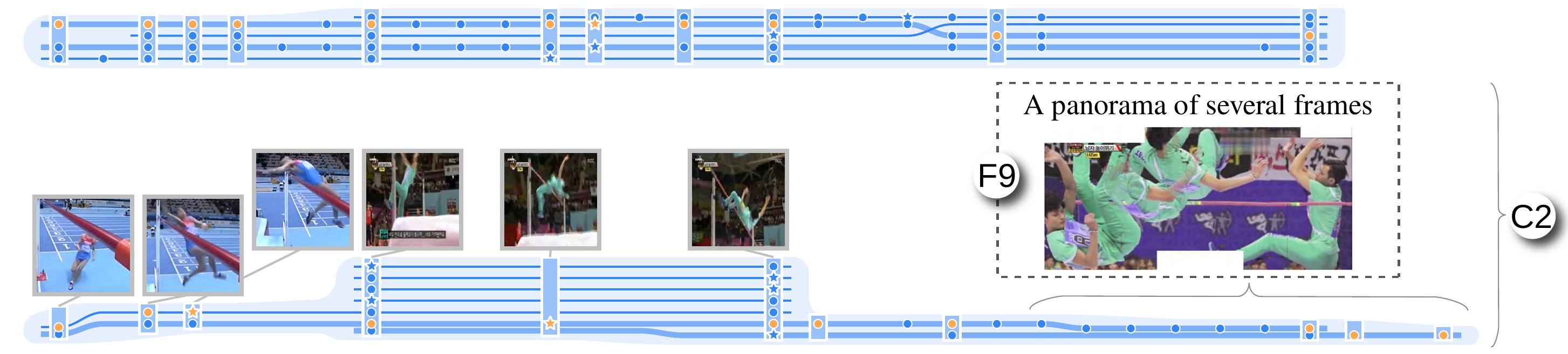}
            \caption{Further analysis on the high jumping category identifies sub-cluster C2 with more frames on the right side.}
        \label{fig:panaroma}
    \vspace{-3mm}
\end{figure*}

\subsection{Case Study}
We conducted a case study with {\Eone} and {\Etwo}, the experts consulted for the requirement analysis, to demonstrate the effectiveness of {\sys} in improving the performance of action localizers trained on single-frame annotations.
The case study was \changjian{conducted} on a subset of \textbf{THUMOS14},
consisting of $1,135$ actions in seven Track and Field sports action categories.
An initial action localizer was trained on the single-frame annotations with SF-Net, which achieved a mAP of \textbf{47.47\%}.
The experts were not satisfied with the performance and wanted to improve it with {\sys}.
As the experts were not involved in the design phase, we introduced the visual design and interactions of {\sys} to them before the case study.
It lasted around 15 minutes.
During the case study, {\Eone} focused on improving the performance of the three categories of jumping sports (high jumping, long jumping, and pole jumping), and {\Etwo} focused on the four categories of throwing sports (javelin throwing, shot put, hammer throwing, and discus throwing).
Here, we take high jumping and javelin throwing as two examples to illustrate the basic idea.  
In the case study, we employed the pair analytics protocol ~\cite{arias2011pair}, 
which allows experts \changjian{to concentrate on analytical tasks while we navigate the tool.}

\subsubsection{High Jumping Action Localization}
{\Eone} started his analysis by examining the uncertainty of the seven action categories (Fig.~\ref{fig:teaser}(a)).
Among them, the high jumping category ($\vcenter{\hbox{\includegraphics[height=1.5\fontcharht\font`\B]{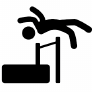}}}$) \changjian{was at the top in the category list, which indicated it had the highest uncertainty.}
To investigate the cause of such high uncertainty, {\Eone} selected this category and switched to the action view.

\begin{figure*}[b]
 \vspace{-3mm}
  \centering
    \includegraphics[width=0.98\linewidth]{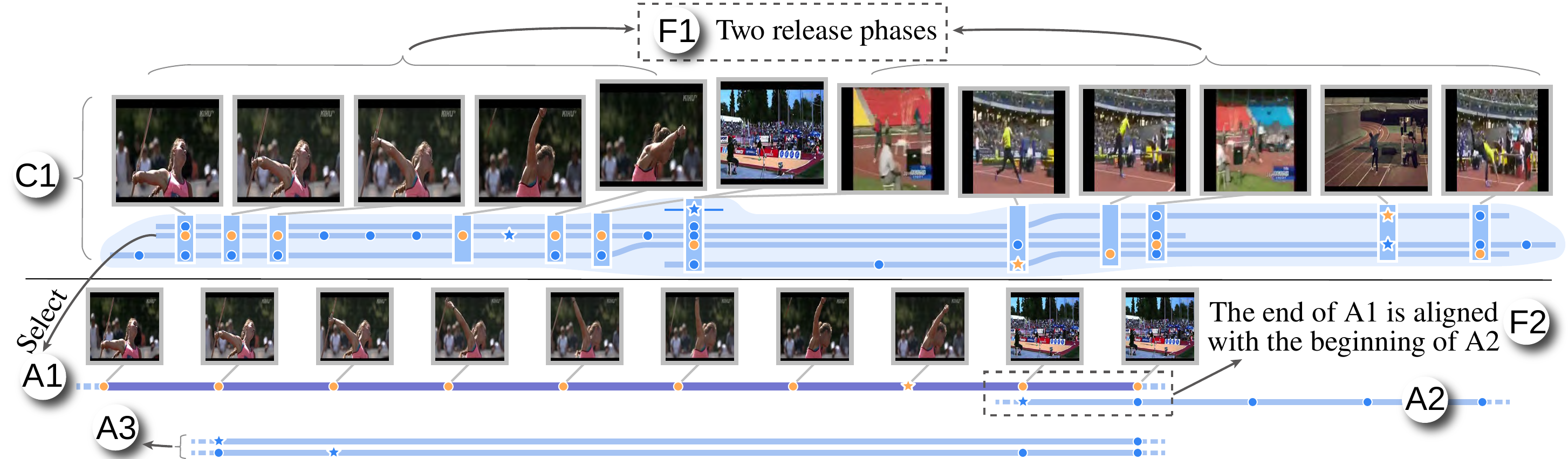}
            \caption{The action view of the javelin throwing category: (a) cluster C1; (b) examination of A1 selected from cluster C1.}
        \label{fig:case2-overview}
\end{figure*}

\myparagraph{Identifying wrong localization results} (\Rexplore).
In the action view, the high jumping actions were clustered into five groups (Fig.~\ref{fig:teaser}(b)).
Typically, a high jumping action starts from an approach phase ($\vcenter{\hbox{\includegraphics[height=1.5\fontcharht\font`\B]{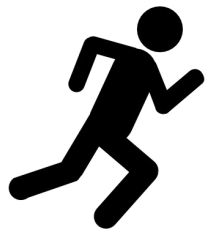}}}$),
followed by a takeoff phase ($\vcenter{\hbox{\includegraphics[height=1.5\fontcharht\font`\B]{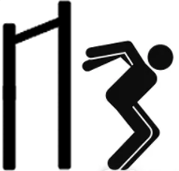}}}$) and a flight phase ($\vcenter{\hbox{\includegraphics[height=1.5\fontcharht\font`\B]{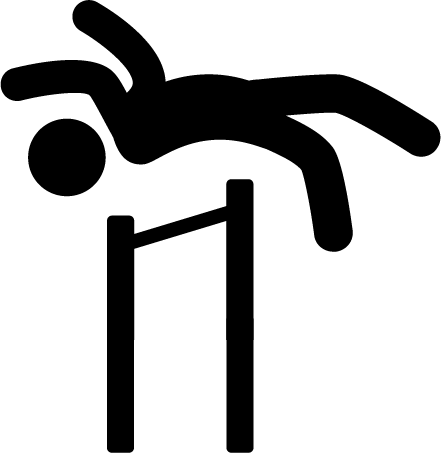}}}$).
By \changjian{checking} the representative frames,
{\Eone} noticed that the starting points of the actions in the top three clusters were not the approach phases but the takeoff phases (Fig.~\ref{fig:teaser}E).
To investigate why the \changjian{starting points} of these clusters were imprecise, he decided to examine them one by one.

\myparagraph{Correcting imprecise boundaries} ({\Rprop}, {\Rstory}).
{\Eone} first selected the top cluster and zoomed in for further analysis.
It contains seven sub-clusters, two of which are shown in
Fig.~\ref{fig:annotating}.
He examined the \changjian{annotated} frames and their locations in these actions and found that
all of them were in the takeoff (Fig.~\ref{fig:annotating}F1) or flight (Fig.~\ref{fig:annotating}F2) phases.
He concluded that the model could not learn to detect the approach phase 
without any \changjian{annotated} frames from this phase.
Thus, {\Eone} decided to add some annotated frames of the approach phase to the training set.
He first analyzed sub-cluster C1 where the left boundaries were the left most of these actions but still imprecise (Fig.~\ref{fig:annotating}C1).
He randomly selected one action (Fig.~\ref{fig:annotating}A1) in this sub-cluster and checked its boundaries.
\changjian{This action was recorded in a TV show. Therefore, the man was not a professional jumper and was using an outdated forward jumping technique.}
{\Eone} found that both the left and right boundaries (frames with orange borders in Fig.~\ref{fig:annotating}F3) were imprecise.
\changjian{As {\Eone} quickly identified the boundaries (frames with green borders in Fig.~\ref{fig:annotating}F3)}, he corrected them directly.
\looseness=-1

The corrected boundaries of A1 were propagated, and 
the visualization was updated accordingly (Fig.~\ref{fig:correcting-alignments}).
In A1's neighbors, more frames of the approach phase were detected (Fig.~\ref{fig:correcting-alignments}F4), and their boundaries were corrected.
This demonstrated the effectiveness of the developed propagation-based action improvement method.

\myparagraph{Correcting misalignments} ({\Rlocate}, {\Rprop}, {\Rstory}).
In the updated visualization, {\Eone} also observed that some frames of four A1's neighbors (A2--A5 in Fig.~\ref{fig:correcting-alignments}) were not aligned with A1.
To investigate the cause, he decided to examine them one by one.
\looseness=-1

{\Eone} first examined the unaligned frames of A2 (Fig.~\ref{fig:correcting-alignments}F5).
Some of these frames belonged to the approach phase but were not aligned with those of A1.
\changjian{Instead, A2's frames left to the approach phase} were aligned  (Fig.~\ref{fig:correcting-alignments}F6).
These frames belonged to the preparation phase and looked very similar to the approach phase.
Such similarity led to the misalignments.
To correct the alignments, {\Eone} decided to align the precise left boundary of A2 with that of A1.
However, \changjian{{\Eone} would need a lot of time to} identify the precise left boundary of A2 because the frames of the preparation and approach phases near the boundary looked similar.
\changjian{To accelerate the annotation process,} 
{\Eone} used the boundary recommendation function.
He first annotated a rough left boundary for this action (Fig.~\ref{fig:correcting-alignments}F7).
Then {\sys} recommended a more precise boundary (Fig.~\ref{fig:correcting-alignments}F8) based on the roughly annotated one.
{\Eone} confirmed that it was the precise boundary and dragged it to be aligned with the left boundary of A1.
Similarly, {\Eone} checked A3, A4, and A5 and corrected their alignments with A1.
These corrections were converted to the must-link constraints and utilized to update other alignments.
\looseness=-1

With the updated alignments, the corrected boundaries were propagated to the \changjian{similar} actions.
A1 and its neighbors were well aligned.
{\Eone} was satisfied with this and proceeded to check the actions in sub-cluster C1.
\changjian{According to the representative frames, the approach phases of these actions were correctly detected.}
This was also verified by the bar chart, which showed that most of the action lengths were longer than before (Fig.~\ref{fig:annotating}B).
He was satisfied with the localization improvement in this sub-cluster and proceeded to examine the remaining sub-clusters.

\myparagraph{Correcting noisy category labels} (\Rlocate).
While examining the remaining sub-clusters, {\Eone} discovered an interesting one, C2 (Fig.~\ref{fig:panaroma}).
Compared to others, the actions in C2 had more frames on the right side.
To investigate this, {\Eone} checked the associated frames of one such action and found that it was a panorama of several frames in a high jumping action (Fig.~\ref{fig:panaroma}F9).
These frames could probably hurt the training process as the optical flow features were used to train the action localizer. 
\changjian{These features represent human motions or human-object interactions in videos because they capture the relative motion between the cameras and the scenes~\cite{carreira2017quo}.}
However, for such a panorama, the optical flow features would capture the camera motion instead of the human motion.
Thus, it should not be regarded as an action (a human motion or a human-object interaction). 
{\Eone} then labeled the associated frames as background frames.

Next, {\Eone} examined the remaining high jumping action clusters in Fig.~\ref{fig:teaser}(b) and corrected misalignments and wrong localization results.
In total, five boundaries and 25 alignments of high jumping actions were corrected.
These corrections were utilized by the propagation-based action improvement method to obtain improved actions, which were used to fine-tune the action localizer.
The mAP was increased from \textbf{47.47\%} to \textbf{47.99\%}.

\subsubsection{Javelin Throwing Action Localization}

After the correction of the high jumping category,
the uncertainty of the javelin throwing category ($\vcenter{\hbox{\includegraphics[height=1.5\fontcharht\font`\B]{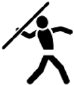}}}$) became the highest.
Therefore, {\Etwo} selected this category for further analysis.

\myparagraph{Identifying wrong localization results} (\Rexplore).
In the action view (Fig.~\ref{fig:case2-overview}), 
the representative frames of cluster C1 revealed the presence of two release phases
$\vcenter{\hbox{\includegraphics[height=1.5\fontcharht\font`\B]{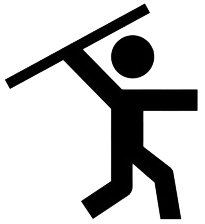}}}$ (Fig.~\ref{fig:case2-overview}F1).
It is abnormal since the actions in a cluster should be aligned without any repetitive phases.
This phenomenon indicated that some actions in this cluster were not aligned properly.
To investigate the cause of such misalignments, {\Etwo} selected one action, A1, for further examination.

\myparagraph{Correcting misalignments} ({\Rprop}, {\Rstory}).
By examining the alignments between the selected action A1 and its neighbors, {\Etwo} found that
the end of action A1 was aligned with the beginning of action A2 (Fig.~\ref{fig:case2-overview}F2).
It was abnormal because the beginnings of two actions should be aligned, as well as their ends.
Further investigation revealed that a set of identical frames \changjian{were detected} in both A1 and A2, which caused the abnormal alignments.
These abnormal alignments resulted in incorrect localization results by preventing \changjian{the same action phases being aligned}.
To address this issue, {\Etwo} corrected the alignments directly by dragging and dropping.
Additionally, he examined and corrected the alignments of other neighbors (Fig.~\ref{fig:case2-overview}A3).

{\Etwo} continued to examine and correct the alignments and localization results of other clusters.
In total, five boundaries and 21 alignments of javelin throwing actions were corrected.
The mAP was improved from \textbf{47.99\%} to \textbf{49.56\%}.

\subsubsection{Other Action Localization and Post Analysis}
\myparagraph{Actions in other categories}. Similar to the analysis of the high jumping and the javelin throwing categories,
{\Eone} and {\Etwo} analyzed the remaining five categories.
24 more boundaries and 165 more alignments were corrected.
The mAP was increased from \textbf{49.56\%} to \textbf{58.29\%}.
Overall, the experts improved the mAP of all the seven categories from \textbf{47.47\%} to \textbf{58.29\%} by correcting a total of 34 boundary annotations and 211 alignments.
\looseness=-1

\myparagraph{Post analysis}. After the case study, we conducted a quantitative evaluation to demonstrate the annotation efficiency of {\sys} in comparison with a human-in-the-loop method without ActLocalizer. 
The latter selects the most uncertain actions for users to annotate, which is one of the most widely-used active learning methods to save annotation efforts~\cite{lewis1995sequential}.
We compared the two methods in terms of the number of corrected boundaries and alignments, annotation time, and mAP.
The annotation time for the baseline method was estimated based on the previous study of Ma~\etal~\cite{ma2020sf}, 
\changjian{which indicated an average of five minutes for annotating action boundaries in a one-minute video.}
The results summarized in Table 2 show that despite requiring additional must-link/cannot-link constraints, {\sys} largely reduces the number of boundaries and annotation time, indicating improved annotation efficiency.
For example, {\sys} saves 78\% annotation time compared with the baseline.
Additionally, {\sys} achieves a higher mAP as well.

\setlength\tabcolsep{5pt}
\begin{table}[t]
    \centering
        \caption{
    The numbers of corrected boundaries and alignments, as well as annotation time comparison between {\sys} and the baseline.
    \looseness=-1
    }
        {
    \begin{tabular}{l|c|c|c|c}
     \toprule
     Method  & \tabincell{c}{\# boundaries} & \# alignments & time & mAP \\
      \midrule
      Baseline   & $113$ & 0  & $4.49$ h  & $57.11\%$   \\
     \hline
      {\sys} & $34$ & 211  & $0.99$ h & $58.29\%$ \\
     \bottomrule
    \end{tabular} 
    }
    \label{tab:case-performance} 
    \end{table}
\setlength\tabcolsep{6pt}

%% file: 6-discussion.tex
\section{Expert Feedback and Discussion}
\label{sec:discussion}

Following the case study, four semi-structured interviews were conducted with the four experts we worked with.
As {\Ethree} and {\Efour} did not participate in the case study, we spent around 20 minutes to introduce the tool and the case study before the interviews.
Each interview lasted 30 to 50 minutes.
All the experts provided positive feedback on the usability of {\sys}.
They also pointed out its limitations and suggested directions for future research.

\subsection{Usability}

\myparagraph{Intuitive visual design and simple interactions}.
All the experts appreciated the intuitive design of the action view.
For example, {\Etwo} said,
``Using lines to represent the actions and the distance between them to depict the alignments is very intuitive to me, 
and I quickly became familiar with the design.''
{\Efour} \changjian{appreciated} the simple interactions supported by {\sys}.
\changjian{For example, he found that correcting alignments was intuitive. It involved simply dragging and dropping operations to indicate which ones should or should not be aligned.}
``Since this tool does not require any prior knowledge of the underlying model, I believe that practitioners can learn to use it quickly,'' he commented.

\myparagraph{Reducing annotation efforts}.
All the experts were impressed by a $78\%$ reduction in annotation time.
{\Eone} noted that the boundary recommendation helped save time in annotating imprecise boundaries.
He said, ``When I am unsure about boundaries, I often turn to the boundary recommendation feature.
It largely saves \changjian{my} annotation efforts in the case study.''
{\Etwo} also said that exploring the actions at different levels of detail saved his efforts in identifying wrong localization results,
``It helps me quickly identify wrong localization results at the global level and zoom in for more in-depth analysis.''

\myparagraph{Generalization to non-single-frame-oriented localization methods}.
\changjian{This work focuses on improving the performance of single-frame-oriented temporal action localization methods.}
However, the experts indicated that our method could be directly utilized to enhance other temporal action localization methods.
The propagation-based action improvement method only relies on the predictions of the frames and the similarities between them, 
which can be obtained from any action localizer.
Furthermore, the storyline aims to present the localization results and the alignments between actions, which are independent of the underlying localization methods.

\subsection{Limitation}
\myparagraph{Algorithm scalability}.
The time complexity of the propagation-based action improvement method is $O(m^2)$, where $m$ is the total number of frames in the training set.
For the \textbf{THUMOS14} dataset with 63,575 frames, the localization results can be updated within one second upon the user corrections.
However, when the number of frames reaches millions, the update could take several seconds or even minutes. 
Therefore, it is worth investigating how to accelerate the propagation-based action improvement method in the future.
One potential solution is to develop an incremental algorithm that only updates the prediction of frames affected by the user corrections. 

\myparagraph{Non-utilization of multi-modality data}.
In the current implementation, the action localizer only utilizes image frames.
However, videos also contain other modalities, such as audio.
\changjian{They} can provide complementary information to improve the localization performance. 
For example, 
audio can indicate that two men in a video are conversing even when they have their backs toward the camera.
As a result, it would be valuable to study how to integrate multi-modality data into the localization model and utilize complementary information between different modalities to improve performance.
\changjian{Furthermore, in cases where the performance is not satisfactory, it would be beneficial to explore ways of visually illustrating the complementary relationships between modalities. This can assist users in correcting misalignments and localization errors in a more efficient manner.}

%% file: 7-conclusion.tex
\section{Conclusion}
\label{sec:conclusion}

In this paper, we introduce {\sys}, a visual analysis method to improve the performance of action localizers trained on single-frame annotations. 
The key feature of our method is the tight integration of the propagation-based action improvement method with the storyline visualization to facilitate users in interactively correcting misalignments and wrong localization results.
The corrected alignments are converted to constraints to derive better alignments for other actions. 
Based on the improved alignments, the corrected localization results are propagated to \changjian{similar} actions to further enhance localization performance.
The effectiveness of {\sys} is demonstrated by the reduced annotation efforts in the quantitative evaluation, the presented findings in the case study, and the positive feedback from the experts after the case study.